\documentclass[prb,aps,twocolumn,superscriptaddress,showpacs,nofootinbib]{revtex4-2}

\usepackage{graphicx}
\DeclareUnicodeCharacter{2009}{\,} 
\DeclareUnicodeCharacter{2212}{--}
\DeclareUnicodeCharacter{2A7D}{$\leq$}
\DeclareUnicodeCharacter{03B2}{$\beta$}
\usepackage{dcolumn}
\usepackage{bm}
\usepackage[mathlines]{lineno}
\usepackage{booktabs} 
\usepackage{xcolor}

\usepackage[utf8]{inputenc}
\usepackage[T1]{fontenc}
\usepackage{mathptmx}
\usepackage{etoolbox}

\begin{document}

 \title{High-pressure floating zone crystal growth of Sr$_2$IrO$_4$}

\author{Steven J. Gomez Alvarado}
\altaffiliation{These authors contributed equally to this work.}
\affiliation{Materials Department, University of California, Santa Barbara, CA, USA}
\author{Yiming Pang}
\altaffiliation{These authors contributed equally to this work.}
\affiliation{Materials Department, University of California, Santa Barbara, CA, USA}
\author{Pedro A. Barrera}
\affiliation{Materials Department, University of California, Santa Barbara, CA, USA}
\author{Dibyata Rout}
\affiliation{Materials Department, University of California, Santa Barbara, CA, USA}
\author{Claudia Robison}
\affiliation{Materials Department, University of California, Santa Barbara, CA, USA}
\author{Zach Porter}
\affiliation{Materials Department, University of California, Santa Barbara, CA, USA}
\author{Hanna Z. Porter}
\affiliation{Materials Department, University of California, Santa Barbara, CA, USA}
\author{Erick A. Lawrence}
\affiliation{Materials Department, University of California, Santa Barbara, CA, USA}
\author{Euan N. Bassey}
\affiliation{Materials Department, University of California, Santa Barbara, CA, USA}

\author{Stephen D. Wilson}
\altaffiliation{stephendwilson@ucsb.edu}
\affiliation{Materials Department, University of California, Santa Barbara, CA, USA}

\date{\today}

\begin{abstract}
Here we demonstrate the floating zone crystal growth of the $J_\mathrm{eff}=1/2$ Mott insulator Sr$_2$IrO$_4$. Historically, the growth of iridates from a ternary melt has been precluded by the extreme vapor pressure of the metal oxide species and the difficulty of maintaining the correct oxidation state of Ir at high temperatures. Here, we show that the application of a high-pressure oxygen growth environment stabilizes the Sr$_2$IrO$_4$ phase, leading to the first demonstration of cm$^{3}$-scale crystals. In contrast to the conventional SrCl$_2$ flux growth method, where poor control over disorder leads to strong sample dependence, the high-pressure floating zone growth enables active control over the homogeneity of the melt. Crystals grown via this technique possess qualitatively similar properties to those grown via flux, with a relatively sharp onset of antiferromagnetic order observed in temperature-dependent magnetization. Further, we demonstrate that by tuning the mixing rate of the melt, we are able to grow natively hole-doped Sr$_2$Ir$_{1-y}$O$_4$, which exhibits a strongly modified magnetic and electronic response.

\end{abstract}

\maketitle
\section{Introduction}

\begin{figure*}[tb!]
	\centering
	\includegraphics[width=1.0\linewidth]{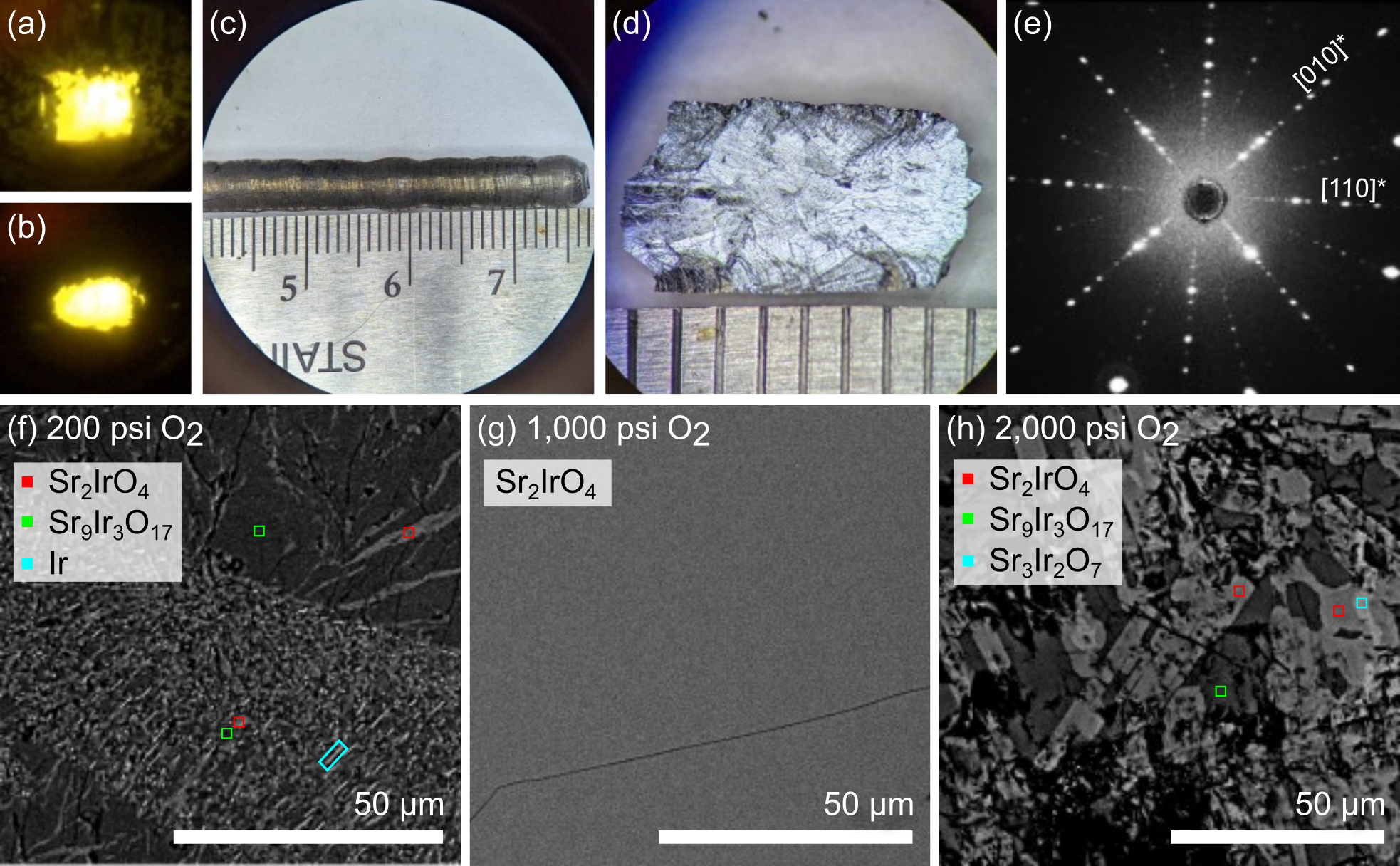}
	\caption{\label{fig:growth_figure}(a) Images captured during the growth upon joining and (b) 15 minutes into the growth. (c) Optical microscope image of the boule as-grown and (d) after mechanical cleavage. (e) Laue diffraction image from the cleaved 001 surface. (f) Scanning electron micrographs of polished cross sections of the boule for several growths at oxygen pressures of 200~psi (14~bar), (g) 1,000~psi (70~bar), and (h) 2,000~psi (140~bar). {Panel (g) was collected on a sample of nearly phase-pure Sr$_2$IrO$_4$,} and corresponding compositional analyses for panels (f-h) are provided in Figure S1 and Table SI.}
\end{figure*}

\begin{figure}
	\centering
	\includegraphics[width=1.0\linewidth]{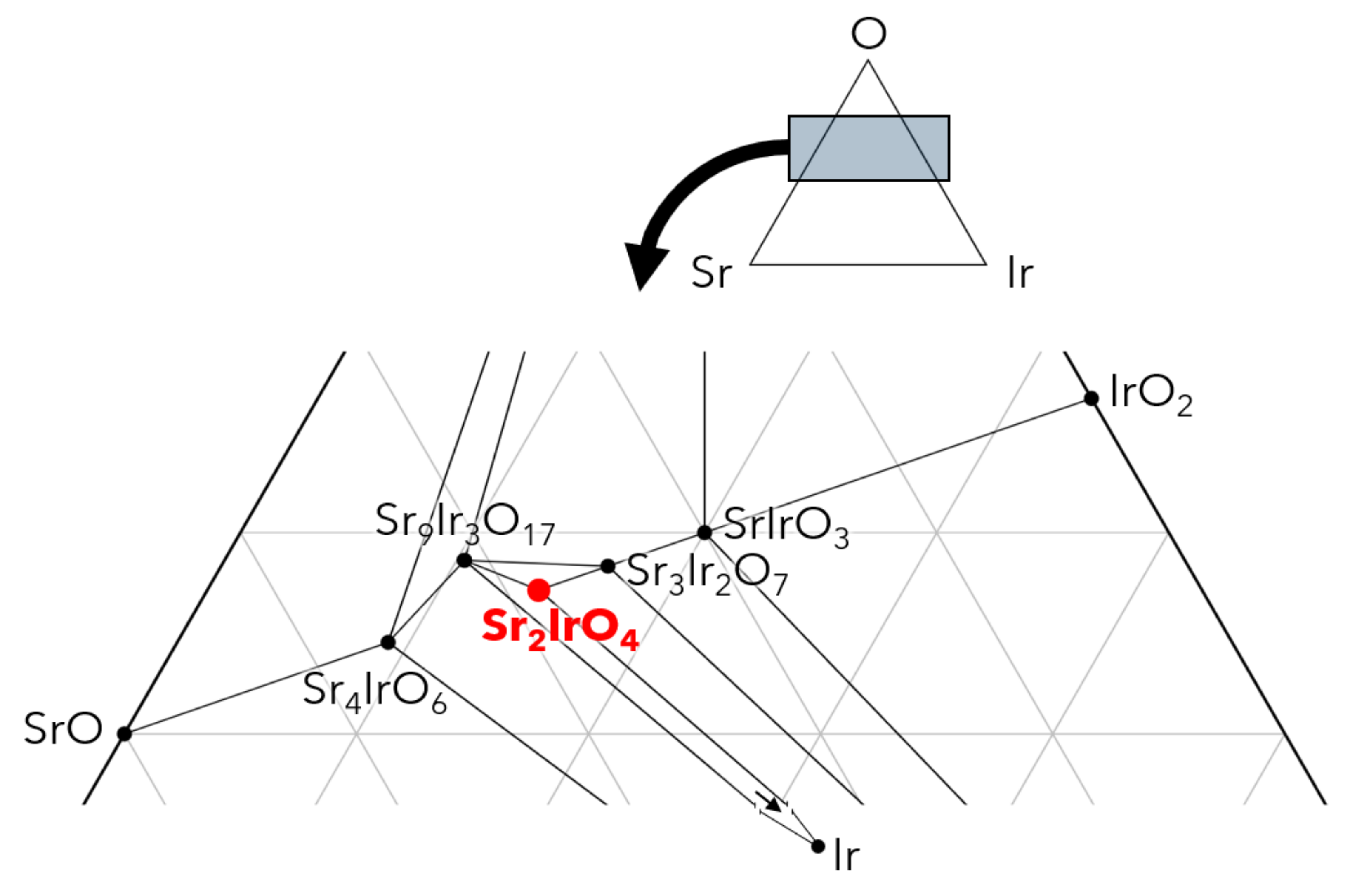}
	\caption{\label{fig:sriro_phase_diagram}The empirical Sr-Ir-O ternary phase diagram obtained from growth trials in the high-pressure laser floating zone furnace near the Sr$_2$IrO$_4$ composition.}
\end{figure}

For over a decade, the $J_\mathrm{eff}=1/2$ Mott insulating state realized in iridium oxides has garnered significant interest owing to its relevance to high-temperature superconductivity, topology, quantum computation, and other applications \cite{jackeli_mott_2009,kim_novel_2008,moon_dimensionality-controlled_2008,kim_phase-sensitive_2009,hwan_chun_direct_2015,takayama_hyperhoneycomb_2015,lu_jeff_2020}.
Within this family of materials, the quasi-2D layered perovskite Sr$_2$IrO$_4$ stands out as a close analogue in crystalline, electronic, and magnetic structure to the Mott insulator La$_2$CuO$_4$ \cite{rau_spin-orbit_2016, pesin_mott_2010, witczak-krempa_correlated_2014, kim_novel_2008, mitchell_sr2iro4_2015, lu_jeff_2020}, which is a parent compound to high-temperature superconductivity upon hole doping \cite{bednorz_possible_1986,uchida_high_1987,takagi_high-tc_1987}.
While theoretical work has indeed predicted superconducting states in Sr$_2$IrO$_4$~at both hole-doped \cite{meng_odd-parity_2014} and electron-doped \cite{wang_twisted_2011,arita_ab_2012,watanabe_monte_2013,yang_superconductivity_2014} compositions, definitive signatures of the superconducting state (\textit{i.e.}, zero resistance and diamagnetism) have thus far remained elusive. 
Despite this, the observation of the related cuprate phenomenology such as Fermi arcs and a pseudogap phase \cite{kim_novel_2008,kim_fermi_2014,yan_electron-doped_2015,de_la_torre_collapse_2015}, as well as the recent observation of a spin density wave \cite{chen_unidirectional_2018} in lightly electron-doped (Sr$_{1-x}$La$_x$)$_2$IrO$_4$,~places this system tantalizingly close to a superconducting state, demonstrating that the analogy to La$_2$CuO$_4$ is robust.

A key limitation in current studies of the magnetic and electronic phase behaviors in bulk crystals of Sr$_2$IrO$_4$ and its derivatives is achieving satisfactory control over nanoscale disorder, homogeneity, and dopant solubility with existing flux-based crystal growth techniques. 
In a typical flux growth, a melt of SrCl$_2$ held within a Pt crucible is super-saturated with a stoichiometric Sr:Ir ratio of 2:1 and cooled through the melting point to precipitate small ($<$mm$^3$-scale) plate-like crystals.
Several recent studies have proposed that the $J_\mathrm{eff}=1/2$ Mott state in Sr$_2$IrO$_4$ is highly sensitive to processing conditions (\textit{e.g.}, heating profile, crucible material, and applied magnetic field) as demonstrated by a notable sample dependence of the magnetic and electronic properties in flux-grown crystals \cite{sung_crystal_2016,cao_quest_2020,kim_single_2022,pellatz_magnetosynthesis_2023}. 
In electron-doped (Sr$_{1-x}$La$_x$)$_2$IrO$_4$, control over the La$^{3+}$ dopant solubility, stoichiometry, and homogeneity in flux-grown crystals remains poor, and the Mott insulating state persists up to maximal solubilities near $x\approx0.06$ \cite{chen_influence_2015, gretarsson_persistent_2016}. 
Scanning tunneling microscopy measurements have revealed nanoscale phase segregation of metallic and insulating puddles in doped samples \cite{chen_influence_2015,battisti_universality_2017,sun_evidence_2021}, indicating that electronic transport remains percolative in nature and is not reflective of a global metal.
Additionally, the small sample volumes of flux-grown crystals complicate the use of weakly interacting probes such as neutron scattering, which is of particular relevance in the study of the $J_\mathrm{eff}=1/2$ ground state and its excitations.

Given these issues, there is a clear need to develop new synthesis routes which offer active control over disorder in order to understand its relationship with the observed physical properties.
One possible solution is to employ the floating zone crystal growth method, which allows for well-defined control over processing parameters such as growth rate, mixing rate, heating profile, and chemical potential. The floating zone technique has been employed widely for a range of transition metal oxides, including 4d systems such as such as Sr$_2$RuO$_4$ \cite{kikugawa_single-crystal_2021} and Sr$_2$RhO$_4$ \cite{perry_sr_2006}, but to date has not been successfully applied to any iridium oxides. This is because the growth of 4d and 5d transition metals such as Ru, Rh, and Ir from a molten phase is dramatically complicated by the tendency of the binary transition metal oxides to volatilize as a vapor phase during the growth \cite{chaston_oxidation_1975,kikugawa_single-crystal_2021, perry_sr_2006}. For Ru and Ir in particular, this occurs via a reaction of (Ru/Ir)O$_2$ in the presence of oxygen to form the vapor phase molecule (Ru/Ir)O$_3$. On the other hand, the congruent melting of iridium oxides demands a high oxygen pressure because of the tendency of IrO$_2$ to decompose into Ir metal and O$_2$ at elevated temperatures. Further, the higher melting temperature of the iridates precludes the use of conventional halogen bulbs as heating elements and exacerbates both of these chemical instabilities.

In the present work, we mitigate these issues using a custom-built high-pressure laser floating zone furnace \cite{schmehr_high-pressure_2019, gomez_alvarado_advances_2024}. Relative to traditional floating zone furnaces based on halogen lamps paired with elliptical focusing mirrors, this laser-based furnace allows for finer control over the heating profile in the molten zone and makes accessible a larger range of temperatures. The furnace is capable of applying total gas pressures up to 15,000~psi (1,000~bar) to suppress material volatility, including partial pressures of oxygen of up to 3,000 psi (200~bar) which can be used to tune chemical potential. We demonstrate that this method allows for the first successful growth of cm$^3$-scale single crystals of Sr$_2$IrO$_4$ and (Sr$_{1-x}$La$_x$)$_2$IrO$_4$ stabilized by moderately high oxygen partial pressures combined with relatively fast growth rates and mixing rates. In addition, we show that different processing regimes result in structural changes at the nano- and micro-scale and allow for tuning of the electronic and magnetic responses.

\section{Experimental methods}

\subsection{Polycrystalline synthesis}

Polycrystalline feed and seed rods were prepared by conventional solid state reaction using SrCO$_3$ (99.99\% excluding Ca or Ba, Thermo Fisher Scientific), La$_2$O$_3$ (99.99\% metals basis, VWR International), and IrO$_2$ (99.99\% metals basis, Thermo Fisher Scientific) precursors. In order to compensate for the loss of Ir as IrO$_{3(g)}$ during the floating zone growth, the feed rods were prepared with an excess of Ir in the stoichiometry (Sr$_{1-x}$La$_x$)$_2$Ir$_{1+\gamma}$O$_4$ over the range $0\leq\gamma\leq0.25$. These reagents were ground and pressed into a pellet using a cold isostatic press under a hydrostatic pressure of $P=47,500$~psi (3275~bar). Samples were then reacted in air at 1100~$^\circ$C for 18~hours. Powder X-ray diffraction collected on a ground piece of the reacted pellet reveals the monoclinic SrIrO$_3$ phase \cite{longo_structure_1971} as the only secondary phase, consistent with the excess Ir stoichiometry. Pellets were then reground and pressed into a rod geometry with diameter ranging from 4~mm to 6~mm, and sintered at 1380~$^\circ$C for 12~hours. Powder X-ray diffraction on a ground piece of the sintered rod revealed the formation of a small amount of IrO$_2$ ($\leq2$ wt\%).

\subsection{High-pressure floating zone crystal growth}

Floating zone growth was carried out in a custom laser-based high-pressure furnace \cite{schmehr_high-pressure_2019, gomez_alvarado_advances_2024}. Oxygen partial pressures ranged from $500\leq P_\mathrm{O2} \leq 2,000$~psi (140~bar) and total pressures including an Ar overpressure ranged from $1,000 \leq P \leq 10,000$~psi ($70 \leq P \leq 700$~bar). The volatilization of some Ir as IrO$_{3(g)}$ during growth was unavoidable, and resulted in the continuous deposition of solid IrO$_2$ on the interior of the growth chamber over time as shown in Figure \ref{fig:growth_figure}(a,b). Fortunately, the incident laser beam mitigated deposition on small areas of the quartz window surrounding the molten zone, allowing for continuous viewing of the zone even after several hours of growth.

Growth rates were varied between 80~mm/hr and 200~mm/hr, as slower growth rates resulted in prohibitive losses of Ir. The rotation rates of each rod were varied between 1~rpm and 50~rpm in a counter-rotating configuration, where higher rotation rates were found to dramatically improve steady-state behavior and, thus, control over the Ir stoichiometry in the resulting boule. The incident laser spot was focused using a planoconvex cylindrical lens to an ellipse with principal axes of 1.5~mm~$\times$~5~mm, and the nominal laser power ranged from 280~W to 840~W. The necessary applied power scaled with the diameter of the melt and with the increased convection and strong optical distortions caused by higher gas pressures.

\subsection{Sample characterization}

Laboratory powder X-ray diffraction measurements were carried out using a Panalytical Empyrean diffractometer using Cu--K$\alpha$ radiation. For synchrotron powder X-ray diffraction measurements, powder samples of Sr$_2$IrO$_4$ were sealed in borosilicate capillaries (outer diameter 0.1~mm), and were measured on beamline I11 at the Diamond Light source \cite{tartoni_high-performance_2008,thompson_beamline_2009}. The patterns were collected with an exposure time of 1~minute using a position-sensitive detector (PSD, Mythen2; $\lambda$ = 0.824297~\r{A}) over the range 2.0$^\circ$ $\leq 2\theta \leq$ 65$^\circ$.

Magnetization measurements were carried out using a Quantum Design MPMS3 SQUID magnetometer in VSM mode. Resistivity measurements were carried out on samples with bar-like geometries using a standard four-wire setup in a Quantum Design Dynacool PPMS with the Electrical Transport Option.

\section{Results}

\subsection{Microstructural dependence on oxygen pressure}

A representative Sr$_2$IrO$_4$ boule grown under optimized conditions is shown in Figure \ref{fig:growth_figure}(c). The mm-scale ripples on the surface of the boule correlated with the rotation rate of the seed rod, and the oscillations in the diameter (every 5-6~mm for this boule) correlated with the occasional collapse of IrO$_2$ powder build-up from the surface of the remaining feed rod. Regions of Sr$_2$IrO$_4$ were concentrated at the center of the boule with inclusions of either Ir metal or the bilayer perovskite phase Sr$_3$Ir$_2$O$_7$ in the bulk, depending on the oxygen pressure. Stripe-like inclusions of the metastable Ir$^{5+/6+}$ phase Sr$_9$Ir$_3$O$_{17}$ penetrate approximately 1~mm deep from the surface of the boule, likely due to enhanced reaction with O$_2$ at the surface of the melt. Optimized boules cleave readily perpendicular to the growth direction, revealing shiny flat surfaces which were confirmed by Laue diffraction to be the expected 001 plane [Fig. \ref{fig:growth_figure}(d,e)].

The structural phases observed in the floating zone crystals depended heavily on the applied oxygen pressure as shown in Figure \ref{fig:growth_figure}(f-h). At $P_\mathrm{O2}=200$~psi (14~bar), insufficient Ir oxidation resulted in decomposition of the melt into Sr$_2$IrO$_4$, Sr$_9$Ir$_3$O$_{17}$, and Ir metal lamellae. At the slightly higher $P_\mathrm{O2}=500$~psi, much more of the Ir metal was oxidized, thus decreasing the fraction of the Ir-poor Sr$_9$Ir$_3$O$_{17}$ phase and increasing the fraction of the Sr$_2$IrO$_4$ phase. A pressure of $P_\mathrm{O2}=1,000$~psi (70~bar) was found to be sufficient to fully oxidize the Ir and avoid metal inclusions. Any local enrichment of the Ir stoichiometry beyond the Sr$_2$IrO$_4$ composition formed inclusions of Sr$_3$Ir$_2$O$_7$ at these pressures. At a pressure of $P_\mathrm{O2}=2,000$~psi (140~bar), evaporative losses of Ir were exacerbated by increased formation of IrO$_{3(g)}$, and the phase fraction of the Sr$_9$Ir$_3$O$_{17}$ phase in the bulk increased dramatically at the expense of the Sr$_2$IrO$_4$ phase. These results are consistent with the phase diagram presented in Figure \ref{fig:sriro_phase_diagram}, and demonstrate that the applied oxygen pressure strongly controls the resulting phase equilibria.

\begin{figure}[tb!]
	\centering
	\includegraphics[width=1.0\linewidth]{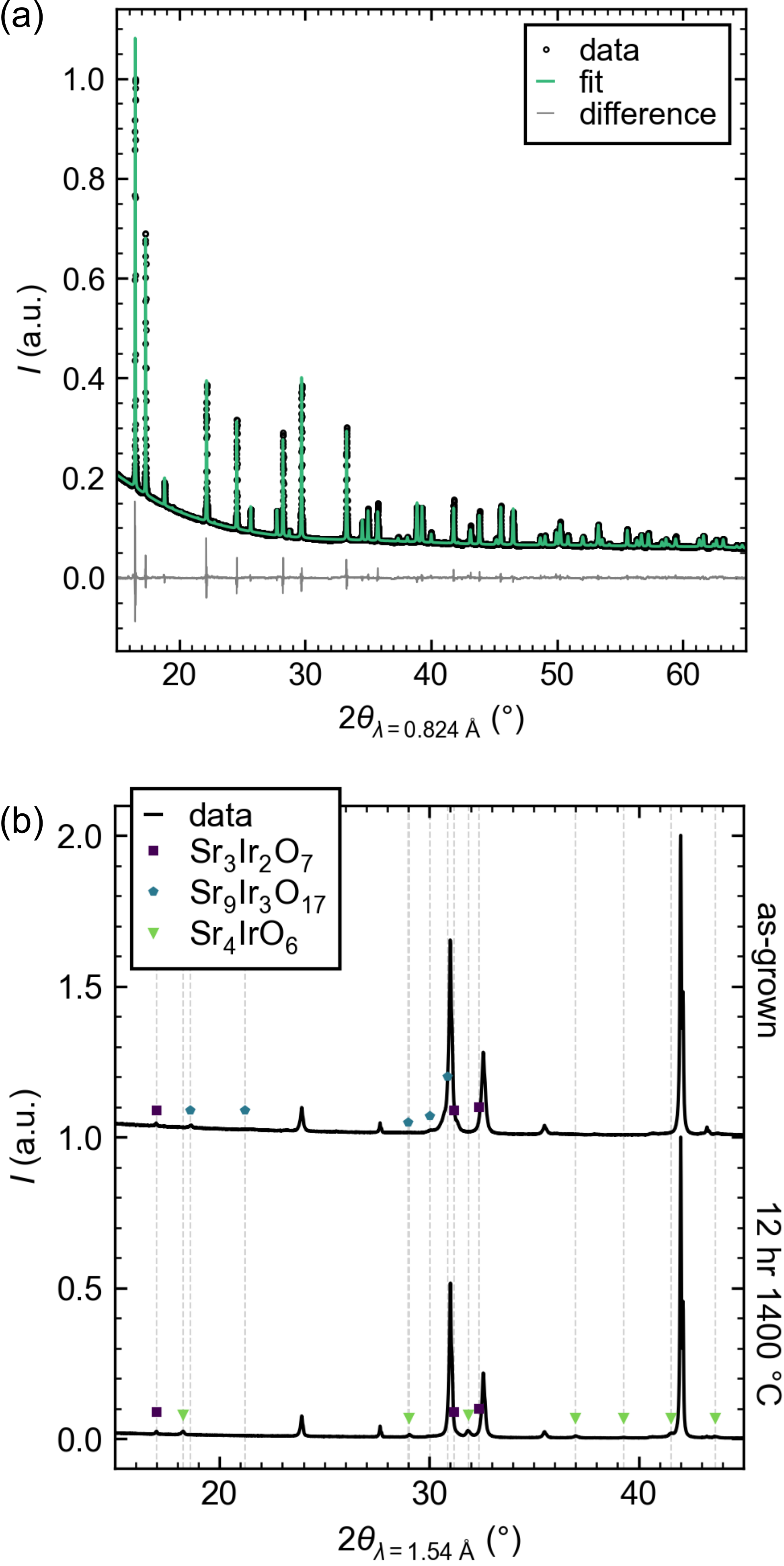}
	\caption{\label{fig:rietveld}(a) Synchrotron powder X-ray diffraction obtained from a ground {phase-pure crystal cleaved from the center of the boule (sample 92A, see Table \ref{tab:crystal_growth})}, including profile fit and difference curves obtained from Rietveld refinement. (b) Laboratory powder X-ray diffraction collected on {ground crystals from the boule including inclusions of secondary phases (sample 92B), both as-grown (top) and after a 12~hour anneal at 1400~$^\circ$C (bottom)}. Note that this sample was not included as a representative growth trial in this study. {The unmarked peaks belong to the Sr$_2$IrO$_4$ phase. Dashed grey lines mark the 2$\theta$ positions of the impurity phases as a guide to the eye for ease of comparison between the two patterns.}}
\end{figure}

{For the remainder of the present analysis, we focus on samples which were grown under the optimized 1,000~psi (70~bar) oxygen partial pressure.} Powder X-ray diffraction measurements on a ground piece of the as-grown boules fit well to a $I4_1/acd$ model of Sr$_2$IrO$_4$, with varying amounts of Sr$_3$Ir$_2$O$_7$ inclusions in the bulk and Sr$_9$Ir$_3$O$_{17}$ confined to the edges of the boule [Fig. \ref{fig:rietveld}(a)]. {We note that we are not able to distinguish between the lower-symmetry $I4_1/a$ model reported for crystals grown in Pt crucibles \cite{torchinsky_structural_2015,dhital_neutron_2013,ye_magnetic_2013,ye_structure_2015} and the $I4_1/acd$ model recently reported for crystals grown in Ir crucibles \cite{kim_single_2022}.}
Post-growth annealing at 1400~$^\circ$C for several hours causes the metastable Sr$_9$Ir$_3$O$_{17}$ phase at the edges of the boule to decompose into Sr$_4$IrO$_6$ and Sr$_2$IrO$_4$, without significantly enhancing the phase fraction of Sr$_2$IrO$_4$ [Fig. \ref{fig:rietveld}(b)].

Given the presence of Sr$_3$Ir$_2$O$_7$ inclusions in most samples, a series of samples with various Ir excess stoichiometries ($\gamma$) were grown and characterized as summarized in Table \ref{tab:crystal_growth}. Despite the wide range of input stoichiometries, inclusions of Sr$_3$Ir$_2$O$_7$ persisted at mole fractions near $x_{327} \leq 0.05$ for Ir excess stoichiometries down to $\gamma = 0.02$, below which the Ir content in the molten zone was too low to stabilize steady-state crystallization of the Sr$_2$IrO$_4$ phase in the resulting boule. One possible explanation for the persistent fraction of this Ir-rich impurity is that the significant deposition and build-up of IrO$_2$ powder on the feed rod is carried into the melt during growth and creates local Ir-rich pockets. This was also observed visually during growth, where a visible collapse of IrO$_2$ powder from the surface of the feed rod typically correlated with a region of significant amounts of Sr$_3$Ir$_2$O$_7$ in the boule. Interestingly, the fraction of Sr$_3$Ir$_2$O$_7$ seemed minimal in compositions near $\gamma=0.08$, and we demonstrate later that these samples exhibit the highest magnetic ordering temperatures and in-plane saturated magnetizations.

\subsection{Mixing-induced changes in bulk physical properties}



\begin{table}
	\centering
	\caption{Crystal growth parameters for a set of representative growth trials at $P_\mathrm{O2}$~=~1{,}000~psi (70~bar) with $P_\mathrm{Ar}$~=~4{,}000~psi (280~bar), their measured magnetic ordering temperature (peak in $dM/dT$), and saturated magnetization. The division indicates grouping by fast mixing (above) and slow mixing (below).}
	\label{tab:crystal_growth}
	\scalebox{1.0}{
		\begin{tabular}{@{} l @{\hspace{10pt}} c @{\hspace{10pt}} c @{\hspace{10pt}} c @{\hspace{10pt}} c @{\hspace{10pt}} c @{\hspace{10pt}} c @{}}
			\toprule
			ID & $\gamma$& $\omega_\mathrm{net}$ & $\dot y$   & $x_{327}$ & $T_\mathrm{N}$ & $M_\mathrm{sat}$      \\
               &   (mol) &                 (rpm) & (mm/h)     &    (m/m)  & (K)            & ($\mu_\mathrm{B}$/Ir) \\ 
			\midrule
			102D & 0.02 & 100  & 120 & 0.11 & 197  & 0.065\\
			100D & 0.04 & 100  & 120 & 0.05 & 219  & 0.074\\
			101C & 0.06 & 100  & 120 & 0.00 & 222  & 0.073\\
			97D  & 0.08 & 100  & 120 & 0.02 & 222  & 0.083\\ 
			92A  & 0.10 & 100  & 120 & 0.00 & 222  & 0.082\\
			90C  & 0.18 & 100  & 120 & 0.09 & 206  & 0.060\\
			\midrule
			86H  & 0.02 & 36   & 100 & 0.06 & 141  & 0.026\\
			84B  & 0.08 & 51   & 120 & 0.00 & 153  & 0.026\\
			79D  & 0.14 & 36   & 120 & 0.06 & 181  & 0.039\\
            78A (1.5\% La)  & 0.16 & 36   & 120 & 0.03 & 206  & 0.071\\
			\bottomrule
		\end{tabular}
	}
\end{table}

Our investigation of the bulk physical properties seeks to determine how the electronic and magnetic ground states respond to varying levels of disorder in the grown crystals. 
The electronic ground state of Sr$_2$IrO$_4$ is a Mott insulating state stabilized by the strong spin-orbit energy scale. As in La$_2$CuO$_4$, this Mott state coincides with the formation of an in-plane antiferromagnetic ordered state within the IrO$_6$ perovskite layers with an ordering temperature $T_N\approx225~\mathrm{K}$, though in the case of Sr$_2$IrO$_4$ the strong spin-orbit coupling induces a slight canting of the moments by an angle $\theta \approx 11^\circ$. This leads to a weak ferromagnetic moment within each layer of IrO$_6$ octahedra, and thus a weak ferromagnetic response in the magnetization. In the absence of disorder, secondary interlayer exchange interactions result in a ``$+--+$'' stacking of these net moments along the $c$ axis in the ground state, leading to an ideally net zero moment at low temperature. A study of isoelectronic disorder via Ca$^{2+}$ substitution within the SrO planes disrupts the ``$+--+$'' ordering \cite{chen_structural_2016}, leading to enhanced low-field polarization of the planar moments. Thus, the moment remains finite at base temperature for real samples.

\begin{figure*}[t]
	\centering
	\includegraphics[width=1.0\linewidth]{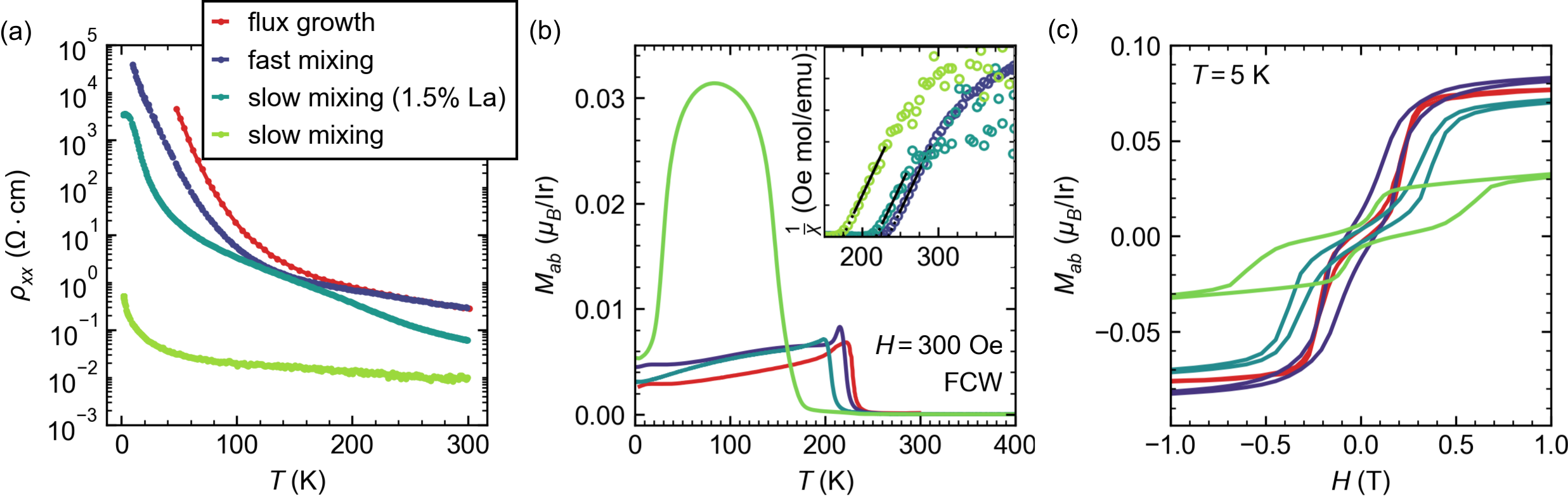}
	\caption{\label{fig:properties}(a) Electrical resistivity vs. temperature (b) $ab$ magnetization vs. temperature with Curie-Weiss fits shown in inset, and (c) $ab$ magnetization vs. field for flux grown,  fast-mixed (sample 97D), and slow-mixed (sample 84B) Sr$_2$IrO$_4$, as well as slow-mixed, lightly electron-doped (Sr$_{0.985}$La$_{0.015}$)$_2$IrO$_4$ (sample 78A).}
\end{figure*}

With the ability to control the homogeneity of the melt via counter-rotation of the feed and seed rods, we demonstrate that samples grown under different mixing conditions exhibit dramatically different magnetic properties as summarized in Table \ref{tab:crystal_growth}. First we describe the behavior of the samples with maximal homogeneity in the melt, here labeled as ``fast mixing'', where the counter-rotation of each rod at 50~rpm generates a total mixing rate $\omega_\mathrm{net}$ of 100~rpm. Electrical resistivity measurements reveal insulating behavior comparable to that observed in flux-grown crystals
[Fig. \ref{fig:properties}(a)] \cite{chen_influence_2015}. In temperature-dependent magnetization measurements, we observe the expected antiferromagnetic ordering transition near $T_\mathrm{N} \approx 220$~K [Fig. \ref{fig:properties}(b)]. The slightly lower resistivity and $T_\mathrm{N}$ in the floating zone crystals may arise from a small fraction of dopants (\textit{i.e.,} a slight off-stoichiometry below the level detectable via energy-dispersive spectroscopy or XRD measurements), though these properties are also expected to be sensitive to other forms of disorder such as isoelectronic substitution \cite{chen_structural_2016}. Despite this, the magnetic ordering transition is sharper that previously reported for flux-grown samples measured at an equivalent magnetic field \cite{chen_influence_2015}, suggesting a globally cleaner magnetic state even in the presence of a weak doping effect.  The likely cause here is a subtle Ir off-stoichiometry due to the evaporative loss of Ir that mimics the Pt impurities discussed by Kim et al. \cite{kim_single_2022}.  Further optimization of the floating zone growth can likely allow this small off-stoichiometry to be mitigated or it may be compensated via La doping.

In contrast, under ``slow mixing'' conditions, where the net rotation rate was closer to 50~rpm, we observe a clear shift towards less insulating behavior [Fig. \ref{fig:properties}(a)], along with a dramatic suppression of the ordering transition down to approximately $T_\mathrm{N} \approx 150$~K [Fig. \ref{fig:properties}(b)]. 
We also find that the low-temperature saturated magnetization is suppressed to about one-third of the expected value, indicating a weaker net moment in each canted AF layer [Fig. \ref{fig:properties}(c)].
Surprisingly, the measured temperature-dependent magnetization in the intermediate range of $50~\mathrm{K}<T<150~\mathrm{K}$ appears to reach the full saturated magnetization ($M_\mathrm{sat} = 0.03~\mu_\mathrm{B}/\mathrm{Ir}$), despite the relatively weak applied field of $H = 300~\mathrm{Oe}$. 
This is potentially because the planar moments are pinned by disorder to a more polarized state during the (Z)FCW measurements.
Further, a second temperature scale appears at $T^{*}=31~\mathrm{K}$ whereupon the magnetization decreases sharply away from the fully saturated value and converges towards a magnetization comparable to that of the sample grown under fast mixing conditions -- suggesting a reentrance of the staggered ``$+--+$'' phase at low temperature.

A Curie-Weiss fit was performed on representative inverse susceptibility datasets for both mixing conditions at temperatures just above $T_\mathrm{N}$ as a parametrization of the inter-layer magnetic response consistent with previous studies \cite{chen_influence_2015}.
For the ``fast mixing'' sample, we found good agreement with reported values for flux-grown samples, exhibiting a Curie-Weiss temperature of $\theta_\mathrm{CW} = 231(3)~\mathrm{K}$ and an effective paramagnetic moment of $\mu_\mathrm{eff} = 0.58(2)~\mu_\mathrm{B}$. For the ``slow mixing'' sample, we find a lower $\theta_\mathrm{CW} = 173(2)~\mathrm{K}$ while the effective moment $\mu_\mathrm{eff} = 0.57(1)~\mu_\mathrm{B}$ remains comparable.  We stress here that this is just an effective parametrization relative to other crystals and that Curie-Weiss fits in this temperature range are generally invalid due to the large in-plane exchange coupling of Sr$_2$IrO$_4$ \cite{kim_magnetic_2012}.

\begin{table}
	\centering
	\caption{\label{tab:diamond} Structural and refinement parameters for Sr$_2$IrO$_4$ at $T$~=~300~K with space group $I4_1/acd$. The atomic positions of Sr, Ir, O(1), and O(2) are $(0, 0, z)$, $(0, 0, 0)$, $(x, 1/2-x, 1/2)$, and $(0, 0, z)$, respectively. Occupancy factors were fixed at 1.0 for both oxygen sites for these refinements.}
	\scalebox{1.0}{
		\begin{tabular}{lll|ll}
			\toprule
			Sr$_2$IrO$_4$ & \multicolumn{2}{c|}{Slow mixing} & \multicolumn{2}{c}{Fast mixing} \\
			\midrule
                ID             & 86H		 & 79D 		& 90C			& 92A \\
			$a$ (\AA)      &5.48847(12)   &5.48804(19)    &5.49014(17)     &5.48951(14)	   \\
			$c$ (\AA)      &25.7780(6)   &25.7797(9)      &25.78498(9)  	&25.7891(6)    \\
				$V$ (\AA$^3$)  &776.52(4)     &776.45(6)      &777.200(4)          	&777.15(4)    \\
			&&&&\\
			Sr & & & & \\
                $\>\>\>\>\>$occ.      &1.00(7)      &1.00(10)       &1.000(5)       &1.00(9) \\
                $\>\>\>\>\>z$         &0.17551(3)   &0.17510(4) 	&0.17595(2) 	&0.17470(3) \\
                
                Ir & & & & \\
                $\>\>\>\>\>$occ.      &0.988(8)     &0.986(11)      &1.000(6)       &0.993(10) \\
			
			O(1) & & & & \\
                $\>\>\>\>\>x$       &0.2984(7)  	 &0.3035(9) 	&0.2960(6)     	&0.3003(8)  \\
			
			O(2) & & & & \\
                $\>\>\>\>\>z$       &0.0790(2)  	 &0.0787(3) 	&0.0834(2) 	   	&0.0793(3)  \\
			
			Ir-O(1)-Ir  & & & & \\
                $\>\>\>\>\>\alpha~(^\circ)$ &158.1(3) 	 &155.8(4)   	&159.2(2)  		&157.3(3) \\
                $\>\>\>\>\>\theta~(^\circ)$ &11.0(2) 	 &12.1(2)   	&10.4(12)  		&11.4(2) \\
			
			&&&&\\
			$R_{wp}$                     &2.42       &1.55           &1.81           &2.36 \\
			\bottomrule
		\end{tabular}
	}
\end{table}

Whereas isoelectronic Ca$^{2+}$ substitution only mildly suppresses the ordering temperature and enhances the saturated magnetization, this suppression of the magnetic ordering and the decrease in the net planar moment is more reminiscent of behavior observed in electron-doped samples \cite{chen_influence_2015}. This suggests that the observed behavior is not simply structural but is likely driven by a change in the electronic state as well.
In contrast to La-doped samples, however, the slow-mixed samples do not exhibit a strong suppression in the temperature-dependent magnetization as expected for light electron doping and instead the in-plane magnetization exceeds that of the normal state at intermediate temperatures.
This could be because disorder in the slow-mixed samples locally disrupts the onset of the ``$+--+$'' state. This is consistent with the distinct shape of the isothermal magnetization which shows an unusual hysteresis indicative of some irreversibility not seen in the fast-mixed sample or in previous reports of Sr$_2$IrO$_4$ [Fig. \ref{fig:properties}(c)]. 
The profile of the temperature-dependent magnetization observed for slow mixing samples is, however, strikingly similar to the recently reported ``field-altered'' Sr$_2$IrO$_4$, where a static field was applied during the flux growth \cite{cao_quest_2020,pellatz_magnetosynthesis_2023}.
This agreement raises the possibility that the mechanism of magnetic alteration may be linked to changes in the mixing of the liquid phase during flux growth, potentially due to magnetic forces acting on the charged species moving in the liquid phase \cite{SERIES1991305}.

The longitudinal resistivity of the ``fast mixing'' samples shows slightly less insulating behavior than flux-grown crystals likely due to percolation through inclusions of secondary phases embedded in the parent Sr$_2$IrO$_4$ matrix [Fig. \ref{fig:properties}(a)]. Consistent with the observed magnetization data, samples grown with slow mixing are significantly less insulating than samples grown with fast mixing, with a resistivity comparable to that of electron-doped samples of (Sr$_{1-x}$La$_x$)$_2$IrO$_4$ with $x=0.02$ \cite{chen_influence_2015}. Fitting the high-temperature resistivity of the fast mixing sample to an activation law $\rho(T) \propto \exp(\Delta/2k_\mathrm{B}T)$ yields an energy gap $\Delta=89~\mathrm{meV}$, which is comparable to the bulk gap $\Delta=108~\mathrm{meV}$ measured for flux-grown crystals (Fig. S2). {We emphasize here that the gap extracted from these fits is not representative of the true Mott gap, but is instead sensitive to in-gap states which may originate from the presence of dopants or defects in the bulk Sr$_2$IrO$_4$ sample used for transport measurements.}

To investigate the nature of the disorder underlying the changes in the physical properties, we carried out high-resolution synchrotron powder X-ray diffraction on ground crystals grown with both fast mixing and slow mixing conditions.
We performed Rietveld refinements of a structural model against the experimental data, and the results are presented in Table \ref{tab:diamond} and Figure S3.
Our analysis demonstrates that the slow mixing samples have a notably smaller in-plane lattice parameter, suggestive of intralayer disorder.
This may arise from atomic vacancies, where cation (Sr/Ir) vacancies would correspond to hole doping and anion (O(1)/O(2)) vacancies would correspond to electron doping.
An alternative explanation would be a relaxed Ir-O(1)-Ir bond angle in these samples. This would be consistent with the more metallic behavior, owing to a stronger overlap between Ir \textit{d} orbitals, and would also explain the weaker observed magnetization response as the strong spin-orbit coupling causes the $J_\mathrm{eff}=1/2$ moments to couple intimately to the magnitude of the octahedral rotations \cite{boseggia_locking_2013}.
Returning to our analysis of the X-ray data, we find that relaxing the occupancy factors of the oxygen anions leads to an unstable refinement and thus we constrain these at 1.0 for all refinements. Given this constraint, a refinement of the occupancy factors for the cation sites suggests Ir vacancies in the amount of 1\% to 2\% for the slow-mixed samples, suggestive of hole doping, while the Sr site tends to converge to full occupancy.

In order to further verify the above interpretation, we introduced a small fraction of La dopants in the amount of $x=0.015$ (determined via energy-dispersive spectroscopy measurements) during a slow mixing growth, which, in the case of electron (hole) doping should cause further suppression (recovery) of the electrical resistivity and magnetic ordering. We find that this La-doped sample exhibits a stark recovery of the strongly insulating behavior observed for fast mixed samples [Fig. \ref{fig:properties}(a)], strongly supporting the notion of compensation for defect-induced, hole-like carriers. We also find a remarkable recovery of the magnetic order, with a sharp ordering transition observed in the temperature-dependent magnetization and a significant increase in the saturated magnetization [Fig. \ref{fig:properties}(b,c)]. A Curie-Weiss fit to the inverse magnetic susceptibility for this sample yields a Curie-Weiss temperature of $\theta_\mathrm{CW} = 216(4)~\mathrm{K}$ and an effective paramagnetic moment of $\mu_\mathrm{eff} = 0.58(3)~\mu_\mathrm{B}$. Additionally, it is notable that the irreversibility observed in this lightly-doped sample is smaller than both the fast-mixed and the slow-mixed Sr$_2$IrO$_4$, which may indicate light hole doping in the fast-mixed samples as well. Taken together, these findings strongly suggest that slow mixing enables the synthesis of vacancy-doped Sr$_2$Ir$_{1-y}$O$_4$, and that substitution with La$^{3+}$ compensates the holes generated by these vacancies with electrons to induce a recovery of the insulating state and magnetic ordering. If true, then a future growth trial at the same composition of $x=0.015$ under fast mixing conditions should exhibit enhanced suppression of the magnetic ordering relative to its slow-mixed counterpart.

\section{Conclusions}

We have successfully demonstrated the first floating zone crystal growth of an iridate compound via the growth of Sr$_2$IrO$_4$, enabled via high pressure. To our knowledge, these samples are the largest (cm$^3$-scale) crystals currently available. A moderate O$_2$ pressure stabilizes the correct oxidation state, and an inert overpressure of Ar helps mitigate IrO$_3$ volatility during growth. We demonstrate that active control over mixing during the growth allows for tuning of the electronic and magnetic response of the material across a dramatic range of properties, likely due to the formation of Ir vacancies which introduce holes into the system. The successful realization of these large-scale crystals may be expected to allow for deeper investigations of the strongly spin-orbit coupled physics in Sr$_2$IrO$_4$ by weakly interacting probes such as neutron scattering. Further, this new regime of growth conditions may be amenable to controlling both disorder and solubility of dopants which may be an important tuning parameter in the pursuit of unconventional superconductivity in this member of the iridate family of compounds. Lastly, we stress that high-pressure oxygen appears to be an important tuning parameter here, and hope that these breakthroughs translate to other iridium oxide-based compounds of interest.

\section{Acknowledgments}

The authors acknowledge various forms of assistance from Eli Zoghlin, Brenden R. Ortiz, Linus Kautzsch, Sarah Schwarz, Jude Quirinale, Bin Gao, Karthik Rao, and Casandra J. Gomez Alvarado. Additional thanks are given to the staff scientists at beamline I11 of the Diamond Light Source for synchrotron data using block allocation group time under proposal CY36397. 
This work was supported via the UC Santa Barbara NSF Quantum Foundry funded via the Q-AMASE-i program under award DMR-1906325. S.J.G.A. acknowledges additional financial support from the National Science Foundation Graduate Research Fellowship under Grant No. 1650114. This research also made use of the shared facilities of the NSF Materials Research Science and Engineering Center at UC Santa Barbara, Grant No. DMR-2308708.

\section{Author declarations}
\subsection{Conflict of interest}
The author has no conflicts to disclose.	
\subsection{Author contributions}
\noindent\textbf{Steven J. Gomez Alvarado}: Investigation (lead); Methodology (lead); Validation (lead); Visualization (equal); Writing - original draft (lead). Writing - review \& editing (equal). 
\textbf{Yiming Pang}: Investigation (equal); Methodology (equal); Validation (equal).
\textbf{Pedro A. Barrera}: Investigation (equal); Methodology (equal); Validation (equal).
\textbf{Dibyata Rout}: Investigation (equal); Methodology (equal); Validation (equal).
\textbf{Claudia Robison}: Investigation (equal); Methodology (equal); Validation (equal).
\textbf{Zach Porter}: Investigation (equal); Methodology (equal); Validation (equal).
\textbf{Hanna Z. Porter}: Investigation (equal); Methodology (equal); Validation (equal).
\textbf{Erick A. Lawrence}: Investigation (equal); Methodology (equal); Validation (equal).
\textbf{Euan N. Bassey}: Investigation (equal); Methodology (equal); Validation (equal).
\textbf{Stephen D. Wilson}: Conceptualization (lead); Funding acquisition (lead); Project administration (lead); Resources (lead); Supervision (lead); Writing - review \& editing (equal).

\section{Data availability}
The data that support the findings of this study are available from the corresponding author upon reasonable request.

\bibliography{biblio_214}

\newpage
\onecolumngrid
\newpage
\part*{Supplemental Information}

\renewcommand{\arraystretch}{0.6} 
\renewcommand{\thetable}{S\arabic{table}}
\renewcommand{\thefigure}{S\arabic{figure}}
\setcounter{figure}{0}
\setcounter{table}{0}

\begin{figure}[h]
    \centering
    \includegraphics[width=1.0\linewidth]{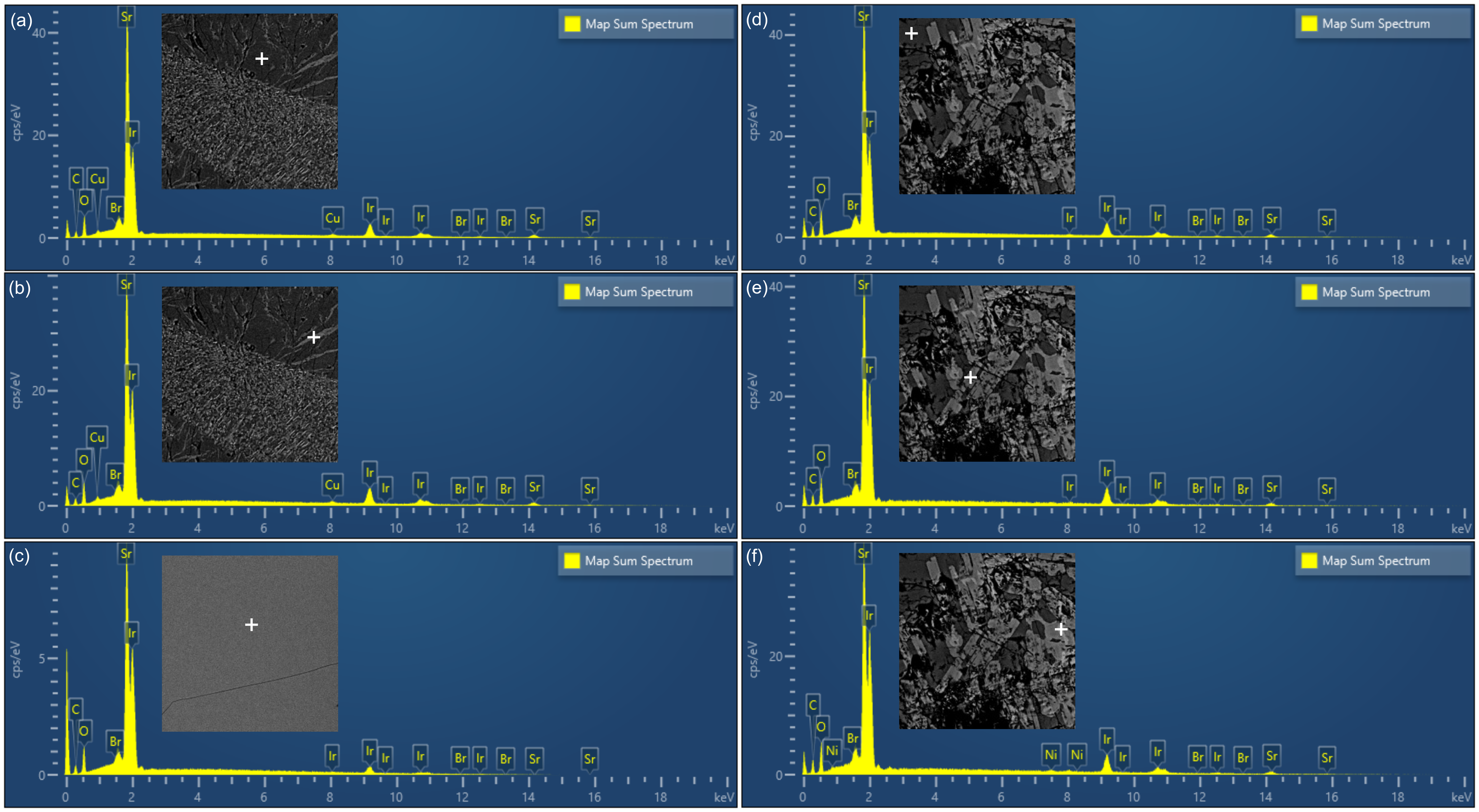}
    \caption{\label{fig:eds9317} Representative energy dispersive spectra associated with the various phases observed during growth. A crosshair marks the probed region in the scanning electron micrograph. Spectra were collected with an accelerating voltage of 20 keV.}
\end{figure}

\begin{table}[h]
	\centering
	\caption{Quantitative energy dispersive spectroscopy results corresponding to the spectra in Fig. S2. Spectra were collected with an accelerating voltage of 20 keV.}
	\label{tab:eds}
	\scalebox{0.96}{
		\begin{tabular}{@{} *{6}{l} @{}} 
			\toprule
			Spectrum & Element & Line Type & Weight \% & Atomic \% & Sr:Ir (m/m)\\  
			\midrule
			(a)   & Sr & L series       & $39.07\pm0.28$   & 16.89 & 2.64 \\
			   & Ir & L series       & $32.49\pm0.33$   & 6.40  &  \\
            (b)   & Sr & L series       & $35.48\pm0.48$   & 16.15 & 2.07 \\
                  & Ir & L series       & $37.66\pm0.62$   & 7.82  &  \\
            (c)   & Sr & L series       & $36.15\pm0.46$   & 17.55 & 2.00 \\
                  & Ir & L series       & $39.73\pm0.54$   & 8.79  &  \\
            (d)   & Sr & L series       & $34.46\pm0.24$   & 12.60 & 2.48 \\
                  & Ir & L series       & $30.44\pm0.31$   & 5.07  &  \\
            (e)   & Sr & L series       & $32.07\pm0.70$   & 12.51 & 1.98 \\
                  & Ir & L series       & $35.57\pm0.98$   & 6.32  &  \\
            (f)   & Sr & L series       & $28.86\pm0.55$   & 10.58 & 1.79 \\
                  & Ir & L series       & $35.36\pm0.83$   & 5.91  &  \\
			\bottomrule
		\end{tabular}
	}

\end{table}

\begin{figure}
    \centering
    \includegraphics[width=0.5\linewidth]{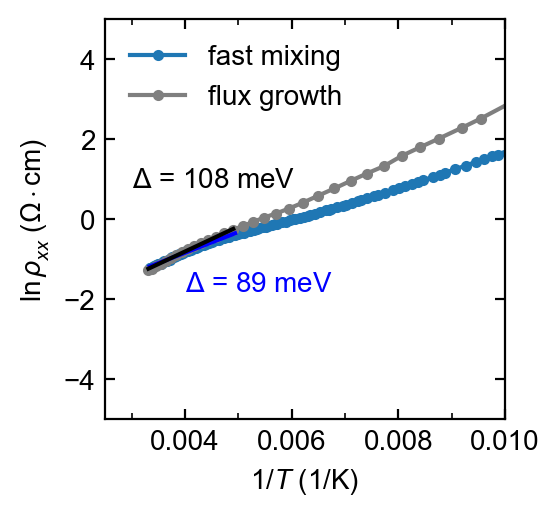}
    \caption{Arrhenius plot of the longitudinal resistivity for the fast mixed sample 92A and a flux grown crystal. Data were fit to the model $\rho(T) \propto\exp(-\Delta/2k_\mathrm{B}T)$.}
    \label{fig:gap}
\end{figure}

\begin{figure}
    \centering
    \includegraphics[width=1.0\linewidth]{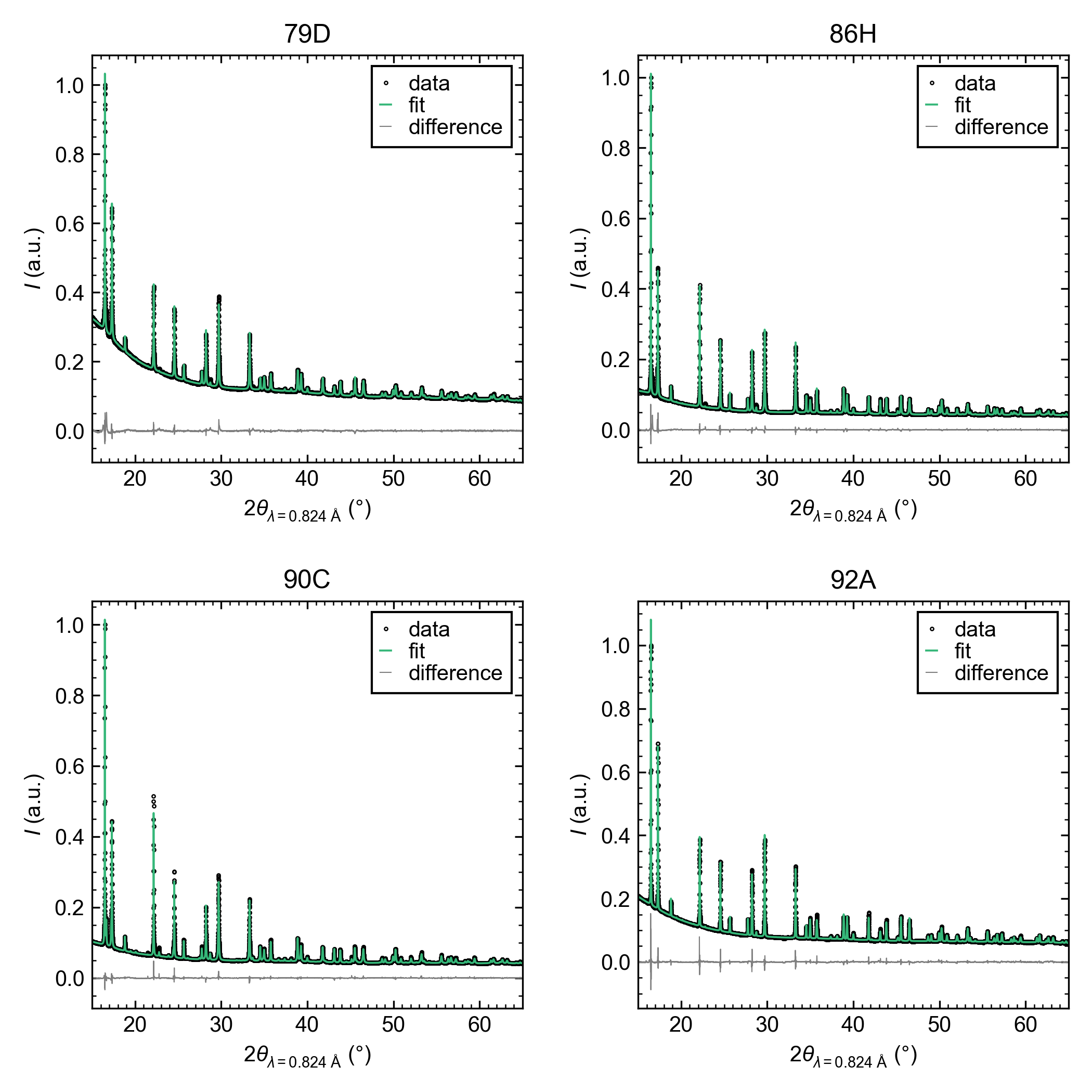}
    \caption{\label{fig:xrd_unshown}Synchrotron powder X-ray diffraction patterns and corresponding Rietveld fit profiles for all samples reported in Table II.}
\end{figure}

\end{document}